\documentclass[aps,prl,showpacs,superscriptaddress,repreint]{revtex4-1}

\usepackage{graphicx}
\usepackage{dcolumn}
\usepackage{bm}
\usepackage{amsmath}
\usepackage{hyperref}
\usepackage{color}
\usepackage{amssymb}
\usepackage{setspace}

\renewcommand{\vec}[1]{{{\bf #1}}}
\makeatother
\usepackage{tikz}

\begin{document}

\begin{spacing}{1.0}

\title{Transition between globule and stretch states of a self-attracting chain in the repulsive active particle bath}

\author{Yi-qi Xia}
\affiliation{Center for Soft Condensed Matter Physics  $\&$ Interdisciplinary Research, College of Physics, Optoelectronics and Energy, Soochow University, Suzhou 215006, China}
\author{Wen-jie Shan}
\affiliation{Center for Soft Condensed Matter Physics  $\&$ Interdisciplinary Research, College of Physics, Optoelectronics and Energy, Soochow University, Suzhou 215006, China}
\author{Wen-de Tian}
\thanks{Email:tianwende@suda.edu.cn}
\affiliation{Center for Soft Condensed Matter Physics  $\&$  Interdisciplinary Research, College of Physics, Optoelectronics and Energy, Soochow University, Suzhou 215006, China}
\author{Kang Chen}
\thanks{Email:kangchen@suda.edu.cn}
\affiliation{Center for Soft Condensed Matter Physics  $\&$  Interdisciplinary Research, College of Physics, Optoelectronics and Energy, Soochow University, Suzhou 215006, China}
\author{Yu-qiang Ma}
\thanks{Email:myqiang@nju.edu.cn}
\affiliation{National Laboratory of Solid State Microstructures and Department of Physics, Nanjing University, Nanjing 210093, China }
\affiliation{Center for Soft Condensed Matter Physics  $\&$  Interdisciplinary Research, College of Physics, Optoelectronics and Energy, Soochow University, Suzhou 215006, China}

\date{\today}

\begin{abstract}
Folding and unfolding of biopolymers are often manipulated in experiment by tuning pH, temperature, single-molecule force or shear field. Here we carry out Brownian dynamics simulations to explore the behavior of a single self-attracting chain in the suspension of self-propelling particles (SPPs). As the propelling force increases, globule-stretch (G-S) transition of the chain happens due to the enhanced disturbance from SPPs. Two distinct mechanisms of the transition in the limits of low and high rotational diffusion rates of SPPs have been observed: shear effect at low rate and collision-induced melting at high rate. The G-S and S-G (stretch-globule) curves form hysteresis loop at low rate, while they merge at high rate. Besides, we find two competing effects result in the non-monotonic dependence of the G-S transition on the SPP density at low rate. Our results suggest an alternative approach to manipulating the folding and unfolding of (bio)polymers by utilizing active agents.
\end{abstract}


\maketitle

Folding and unfolding of bio-macromolecules are ubiquitous phenomena in biological world. For example, incorrect protein folding is the cause of many diseases~\cite{paul06};
DNA condensation and  un-condensation synchronized with cell periods play important roles in cell function~\cite{teif11}. Commonly, many folding and unfolding processes in life are far from equilibrium such as ATP-dependent G-quadruplex unfolding by Bloom helicase~\cite{budhat15}. In experiment, the folding-unfolding transition can be manipulated by tuning pH, temperature, or external force such as by optical tweezers, atomic force microscopy and shear field~\cite{pans06,Norman07,Stephen06}. In polymer physics, analogous phenomena are the transitions of polymer chains between globule and coil/stretch states. In engineering, polymers in shear flow exhibit complex behaviors such as periodic elongation, relaxation, tumbling, and, sometimes, even crystallization~\cite{larson05,doyle00}. Pronounced non-monotonic stretching of polymeric globules was also observed in the external flow field~\cite{katz06}. In recent years, variety of ``colloidal polymers" or ``nanochains" have been fabricated with colloidal particles as the ``monomers" of these chain materials, which makes it possible to study the polymer behaviors at larger length scale~\cite{hill14,chensoftmatter,chenjcp}.

In this letter, we report a new approach to manipulating the G-S and S-G transitions by immersing a self-attracting chain in the repulsive active particle bath. The motion of active particles is inherently non-equilibrium, requiring persistent energy input~\cite{rao13}. A collection of such active particles exhibits novel and intriguing nonequilibrium phenomena such as giant fluctuation, wall accumulation, and motility-induced phase separation~\cite{lowen16}, which hence has attracted growing interests~\cite{sriram10}. Boundary greatly modulate the distribution and motion of the active particles~\cite{Kaiser12,fily14}. Conversely, these active particles exert inhomogeneous pressure on the boundary, dependent on its shape/local curvature~\cite{nikola16}. As a special case,  polymer chains act as a deformable and movable boundary. The cooperation between the passive thermal motion of chain beads and the active non-thermal motion of particles leads to anomalous static and collective dynamic behaviors of the chains/active particles assembly~\cite{Kaiser14,solon15,chen15,smchen,tian15}. 

Unlike the shear flow field which imposes a spatially regular forces on polymer chains, active particles in the suspension exert highly fluctuating and adapting forces on the immersed chains~\cite{lowen16}. It is an open question whether the presence of active particles can trigger the G-S transition of a self-attracting chain and, if yes, how the transition is influenced. By Brownian dynamics simulation in two-dimensional geometry (without hydrodynamic interactions), we find that a first-order-like G-S transition of a self-attracting chain happens with the increase of propelling force on the active particles. Dependent on the rotational diffusion rate $D_r$ of the active particles, the initially collapsed chain can be stretched by the mechanisms of shear effect and/or collision-induced melting. When shear effect dominates, hysteresis loop is formed by the G-S and S-G curves. It originates from the formation of particle layers around the stretched chain which act as kinetic obstacle/barrier for chain collapsing. On the contrary, the layers disappear and the two curves merge together when collision-induced melting leads.

\emph{Model and Simulation Methods}
In our two-dimensional model, an attractive bead-spring chain consisting of $N_p=100$ (passive) beads are mixed with N self-propelling particles(SPPs). The SPPs follow the coupled Langevin equations of translational motion,
$\dot{\vec{r}}_{i}=(F_{a}\hat{n}_{i}(\theta)-\vec{\nabla}_{i}U(\vec{r}))/\gamma+\sqrt{2D_{t}}\vec{\eta}_i(t)$, and rotational motion, $\dot{\theta}_{i}=\sqrt{2D_{r}}\vec{\xi}_i(t)$, in the overdamped limit. $\vec{r}_{i}$ is the position of the $i$th SPP, $\gamma$  the translational friction coefficient. $\hat{n}_{i}(\theta)=(\cos\theta_i,\sin\theta_i)$ is the inherent orientation of the \emph{i}th particle. The translational noise, $\vec{\eta}_i(t)$, and the rotational noise, $\vec{\xi}_i(t)$ are the unit-variance Gaussian white noises, satisfying the fluctuation-dissipation theorem~\cite{weber56}. $U(\vec{r})$ is the pairwise interaction potential; $F_{a}$ is the constant magnitude of the self-propelling force on SPPs; $D_{t}=k_{B}T/\gamma$ and $D_{r}$  are the translational and rotational diffusion constants. The motion of the passive chain beads is solely described by the translational equation with $F_{a}=0$. The bonded harmonic potential between two successive beads of distance $r$ is $U_{b}=\varepsilon_{b}(r-r_{0})^2$, where we set $\varepsilon_{b}=2500k_{B}T/\sigma^{2}$ and $r_{0}=0.98\sigma$. The non-bonded pair interaction is modeled by the smooth-shifted Lennard-Jones potential, $U_{LJ}(r)=4\varepsilon[(\frac{\sigma}{r})^{12}-(\frac{\sigma}{r})^{6}]$ + $\varepsilon_{c}$, where $\varepsilon_{c}$ is the shifted energy that ensure $U_{LJ}(r_{c})$=0. The pair interaction between chain beads is attractive with the cutoff $r_{c}=2.5\sigma$ and the amplitude $\varepsilon_{pp}=3k_{B}T$. The other pair interactions including bead-SPP and SPP-SPP are purely repulsive with $r_{c}=\sqrt[6]{2}\sigma$ and $\varepsilon_{ps}=\varepsilon_{ss}=10k_{B}T$.

We use the home-modified LAMMPS software~\cite{steve95} to perform the simulations. A square box of $100\sigma$$\times$$100\sigma$ with periodic condition in both x and y directions is adopted. Reduced units are used in the simulations by setting $\sigma=1$, $k_{B}T=1$; $\tau=\sigma^{2}/D_{t}$ is the corresponding unit time and $F_{a}$ with the unit of $k_{B}T/\sigma$. We set the friction coefficient $\gamma=10$ and $D_{r}$ as an independent parameter as in the previous work~\cite{tian15}. For each case, it was run by a minimum time of $10^{4}\tau$  with a time step $\Delta t=10^{-4}\tau$.

\emph{Results}
\begin{figure}[t]
  \includegraphics[width=.99\columnwidth,height=.46\columnwidth]{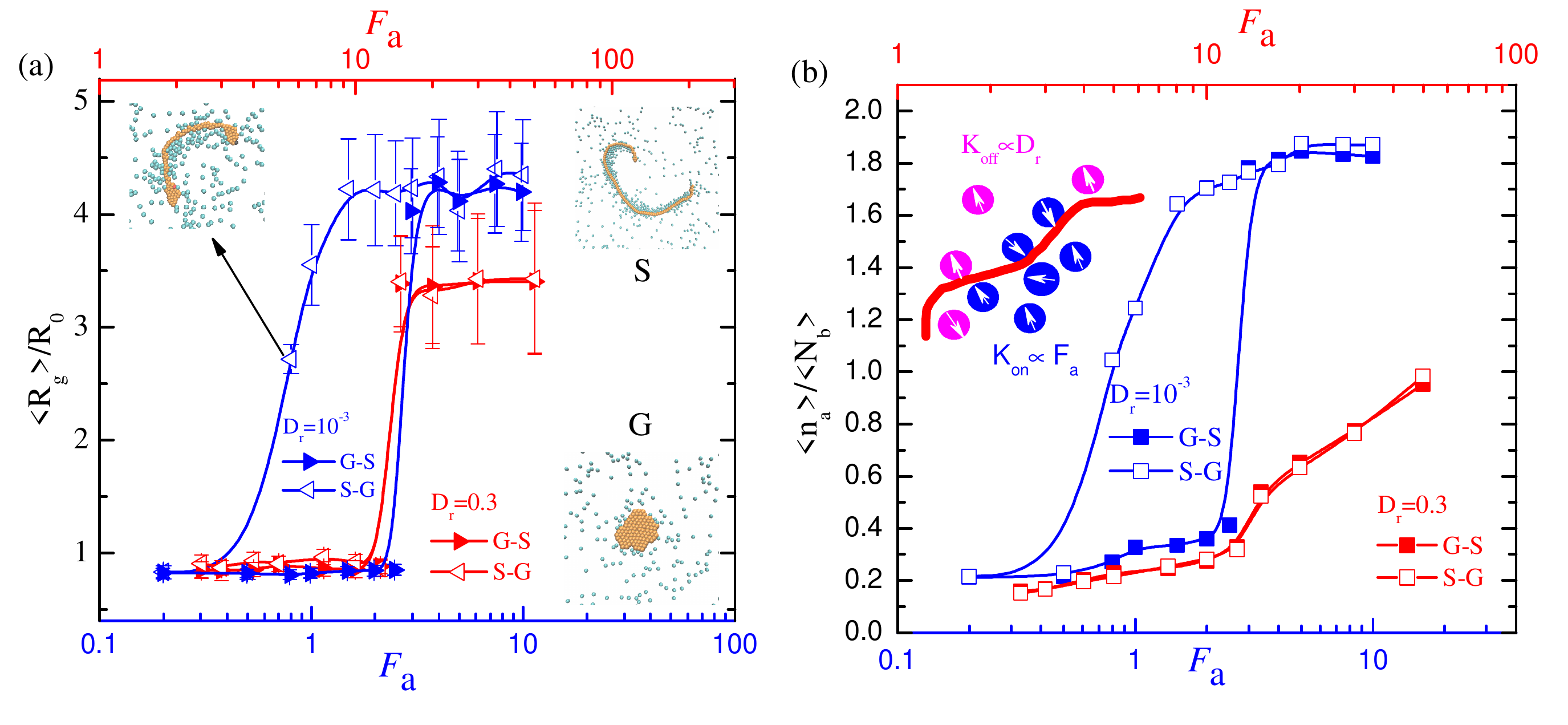}
  \caption{(a) Reduced radius of gyration as a function of propelling force, $F_{a}$, for chain length $N_{p}=100$ and area fraction of particles $\phi=0.05$. The insets are typical snapshots for globule, stretch and intermediate states of the chain. Orange and cyan circles represent chain beads and SPPs, respectively. (b) The corresponding average number of neighboring particles per boundary bead, $\langle n_{a}\rangle/\langle N_{b}\rangle$, as a function of $F_{a}$. The inset is a ketch of SPPs  hitting onto and leaving off the chain.}
   \label{fig:1}
\end{figure}
A collapsed chain is initially prepared by long time simulations in the case of $F_{a}=0$ (passive particles) and particle area fraction $\phi=0.05$. A typical snapshot is shown in the inset of Fig.~\ref{fig:1}a. Starting from these collapsed states, we assign propelling forces of the same magnitude but random directions to the bath particles. As the propelling force increases, a sharp (first-order-like) transition of the chain configuration from collapsed or globule state to stretch state happens at certain critical force $F_{a}^{c}$. Fig.~\ref{fig:1}a shows the transitions in terms of the reduced chain size. $\langle R_{g}\rangle\equiv \langle\sqrt{\sum_j(\vec{r}^{j}-\vec{r}_{com})^{2}}/N_{p}\rangle$ is the mean radius of gyration. $\vec{r}^{j}$ and $\vec{r}_{com}$ are the position vectors of the $j$th bead and the center of mass of the chain, respectively. $R_{0}=0.5\sigma N_{p}^{1/2}$ is the radius of gyration of an ideal Gaussian chain of bond length $\sigma$. $F_{a}^{c}$ depends on another key quantity that characterizes the activity of SPPs, i.e. the rotational diffusion coefficient $D_{r}$, the reciprocal of which gives the persistence time of the propelling force. Larger $D_{r}$ requires larger propelling force for the G-S transition. $F_{a}^{c}$ differs by a factor of 5 for $D_{r}=0.001$ and $D_{r}=0.3$ (Fig.~\ref{fig:1}a). Curiously, the mean chain size in the stretch state depends notably on $D_{r}$ but very weakly on $F_{a}$. This seems to contradict the findings in Ref.~\cite{Kaiser14} that the end-to-end distance of a non-attractive chain in the active bath varies significantly with the propulsion strength.

We then explore the reverse process, i.e. the chain at the stretch state is prepared in the case of large propelling force and then the force is suddenly reduced to a certain lower value. We find that the stretched chain folds back into globule state (i.e. S-G transition) below a threshold $F_{a}^{c'}$. For large $D_{r}=0.3$, the G-S and S-G transition curves coincide, both analogous to a first-order transition. However, it's complicated for small $D_{r}=0.001$. Hysteresis is formed between G-S and S-G transition curves, i.e. S-G transition happens around $F_{a}^{c'} \approx 0.8$ lower than that of the G-S transition. Moreover, this S-G transition is no longer like a first-order transition. Intermediate ``dumbell" states appear, in which the chain is stretched in the middle but with small collapsed lump at each end (inset snapshot of Fig.~\ref{fig:1}(a)). Such intermediate states are kinetically stable during the entire simulation time. The ``stability" has also been tested and verified by several independent long-time runs.

\begin{figure}[t]
  \includegraphics[width=0.99\columnwidth,height=.50\columnwidth]{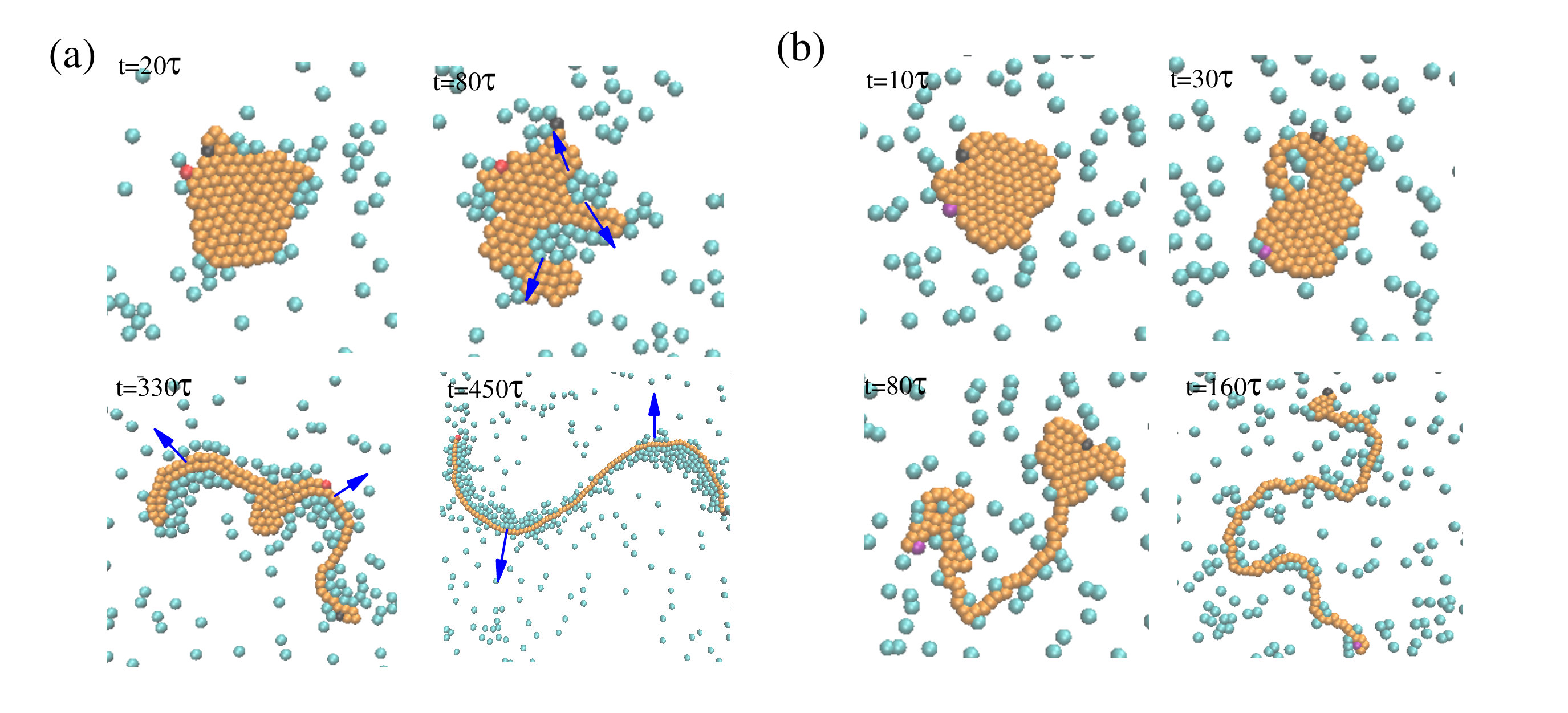}
  \caption{Time evolution of the chain configuration. (a)$F_{a}=5.0, D_{r}=0.001$; blue arrows show the directions of the ''shear" forces. (b) $F_{a}=15.0, D_{r}=0.3$.}
   \label{fig:2}
\end{figure}

Two questions arises: 1) why and how is the collapsed chain stretched by the SPPs; 2) physically, what causes the hysteresis? why does it happen when $D_{r}$ is small while disappear when $D_{r}$ is large? To answer these questions, we examine the folding and unfolding trajectories carefully. The chain acts as a deformable and movable passive boundary and hence the SPPs tend to accumulate around it. The propulsion strength $F_{a}$ (or drift velocity) determines the collision rate, $k_{on}$, between SPPs and the chain, while the persistence time $1/D_{r}$ dictates statistically how long a SPP stick to the chain after collision or the rate that a SPP leaves off the chain, $k_{off}$ (inset of Fig.~\ref{fig:1}b)~\cite{hagan13}. Of course, as aforementioned, the shape of the boundary (local curvature) influences the accumulation of SPPs as well. In Fig.~\ref{fig:1}b, we quantify the degree of aggregation of SPPs around the chain by the ratio $\langle n_{a}\rangle/\langle N_{b}\rangle$. $\langle n_{a}\rangle$ counts the average number of SPPs within the distance of $1.6\sigma$ (the first trough of the SPP-SPP radial distribution function) to the chain. $\langle N_{b}\rangle$ is the average number of beads at the interface. Even though the propulsion strength for $D_{r}=0.3$ is higher than that of $D_{r}=0.001$ in Fig.~\ref{fig:1}b, the degree of particle aggregation for the latter is much larger, especially in the stretch state. Related to this difference, we find two distinct mechanisms of the G-S transition in the limits of large and small $D_{r}$, respectively. In the small $D_{r}$ limit, SPPs keep accumulating strongly around the chain during the unfolding transition (Fig.~\ref{fig:2}a). The forces on the chain by these surrounding SPPs are fluctuating and inhomogeneous and lead to irregular deformation of the collapsed chain lump. In response to the shape deformation, more and more SPPs aggregate in the concave region and generate effective shear force on the chain (see Fig.S1)~\cite{support}. It's this effective shear force, which fully stretch the chain eventually. In contrast, the aggregation of SPPs around the chain in the large $D_{r}$ limit is weak, even the chain is in the stretch state. Notice that, in this limit, the average size of the stretched chain is much smaller than that in the small $D_{r}$ limit (Fig.~\ref{fig:1}a) and moreover, the chain contour is not smooth (not fully stretched) with some local packing of beads (Fig.~\ref{fig:2}b). All these facts indicate that the effective shear force is no longer the major reason of the G-S transition in the large $D_{r}$ limit. Instead, the frequent collisions between the SPPs and the chain at large $D_{r}$ effectively transfers the kinetically energy from the active particles to the chain beads, i.e. the chain is equivalently heated. Beyond a certain propulsion strength, the collapsed chain ''melts" due to the frequent collisions. For the overdamped Brownian motion, we adopt the effective translational diffusion constant $D_{eff}$ (extracted from the MSD curve of chain beads at the normal diffusion regime) as a measure of the effective temperature. For large $D_{r}=0.3$, the ratio of $D_{eff}$ at stretch ($F_{a}=15$) and globule ($F_{a}=0$) states is around 3 (see Fig. S2)~\cite{support}. 

The different physical mechanisms of the G-S transition cause the differences in the reverse S-G process, as well. For large $D_{r}$, the collision strength and rate and hence the effective temperature
of chain beads decrease with the decrease of propelling force. Eventually below certain propelling force the random kinetic energy input from collisions cannot overcome the attractions between beads and the chain undergoes S-G transition. This transition is analogous to the thermodynamic first-order transition and the G-S and S-G curves merge together. On the contrary, the chain is always surrounded by a large amount of SPPs for small $D_{r}$, which become obstacles (effective dynamic barrier) for the folding of the chain. A low enough propelling force is necessary to reduce the stickiness between the chain beads and SPPs, so that the chain can collapse. This is the origin of the hysteresis in Fig.~\ref{fig:1}(a). The dynamic barrier of chain folding is not uniform along the chain. It's weak around the two free ends due to less spatial constraint. And therefore the intermediate states, i.e. the chain with stretched middle part and collapsed two ends are formed in the narrow S-G transition region (inset snapshot in Fig.~\ref{fig:1}a).

\begin{figure}[t]
  \includegraphics[width=.99\columnwidth,height=.46\columnwidth]{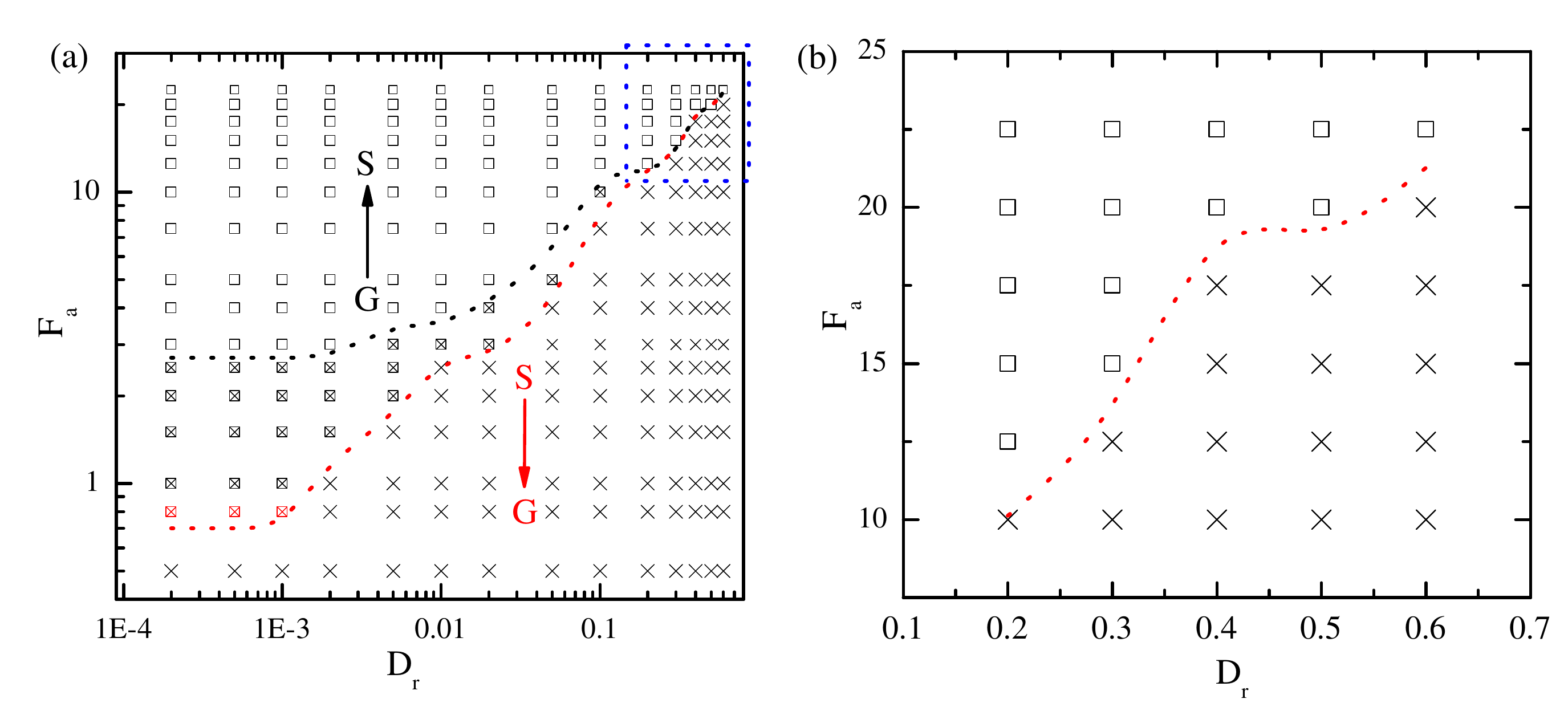}
   \caption{
   (a)Phase diagram of the G-S and S-G transitions in $F_{a}-D_{r}$ space. The black and red dashed lines mark the boundaries of the G($\times$)-S($\Box$) and S-G transitions, respectively. The red mixed symbol of $\times$ and $\Box$ represent the intermediate ''dumbell" states. (b) Enlarged drawing of the phase diagram squared by the blue dashed lines in (a) (large $D_{r}$ regime); here, the red dashed line is the prediction by theory.
   }\label{fig:3}
\end{figure}

$F_a$ and $D_{r}$ are two crucial parameters in determining the active-particle-induced folding and unfolding phenomena. We hence systematically explore the G-S and S-G transitions in the $F_a$-$D_r$ space (Fig.~\ref{fig:3}). As guide for eyes, the dotted black line (upper) and the dotted red line (lower) mark the G-S and S-G transitions, respectively. These two lines are separated when $D_r<0.2$, indicating the appearance of the hysteresis phenomenon. On the contrary, the lines merge together, i.e. the hysteresis disappears when $D_r>0.2$. The hysteresis area gradually decreases in the range $0.001<D_r<0.2$. Notice that these two lines are almost flat when $D_r$ is very small. In this small $D_r$ limit, we give a theoretical interpretation or estimate on $F_{a}^{c}$ with the physics picture that the SPP hitting on the chain exerts $F_a$ on the chain persistently (large persistence time $1/D_{r}$). For the G-S transition, we schematically calculate the minimum $F_a$ that is required for a SPP to overcome the cohesive energy between two non-bonded beads initially in contact (see Fig. S3)~\cite{support}. Appealing to the force and work-energy balance, we obtain $F_{a}^{c} \approx 3$ which agrees well with the simulation result. For the S-G transition, we calculate the maximum $F_a$ under which two adjacent attractive non-bonded beads can push away a SPP (see Fig. S4)~\cite{support}. We obtain $F_{a}^{c'} \approx 0.5$ which is also consistent with the simulation result. As $D_r$ increases, the leaving-off rate $k_{off}$ increases, which means the influence of SPPs on the chain is weakened. Therefore the $F_{a}^{c}$ increases to raise the collision rate $k_{on}$ and enhances the disturbing strength to compensate the negative impact from the increase of $D_r$. In the large $D_r$ limit, the mechanism underlying the G-S and S-G transitions is the collision-induced melting. We make a simple theoretical analysis, for example, on the G-S transition as follows. Averagely, in globule state, $\langle N_{b}\rangle$ boundary beads are in contact with $\langle n_{a}\rangle$ SPPs (Both of them are listed in Table S1~\cite{support}). The SPPs transfer kinetic energy to the passive chain beads through collisions. The kinetic energy of a boundary chain bead can be written as $E_b=\alpha\langle n_{a}\rangle k_BT_{eff}/\langle N_{b}\rangle+k_BT$, where $\alpha$ is the ratio or efficiency of the energy transfer and $k_BT_{eff}=F_a^2/(2\gamma D_r)+k_BT$ is the effective kinetic energy of a SPP\~cite{lowen16}. The transition happens when these boundary beads are ''melted", i.e. $E_b\approx 3k_BT$ (the barrier height of the LJ potential between two non-bonded chain beads). Hence, the transition happens at $F_{a}^{c} \approx\sqrt{2\gamma D_rk_BT(\frac{2\langle N_{b}\rangle}{\alpha\langle n_{a}\rangle}-1)}$. Figure ~\ref{fig:3}(b) shows the transition is well predicted by the equation with fitting parameter $\alpha\approx 0.3$.

\begin{figure}[t]
  \includegraphics[width=.99\columnwidth,height=.46\columnwidth]{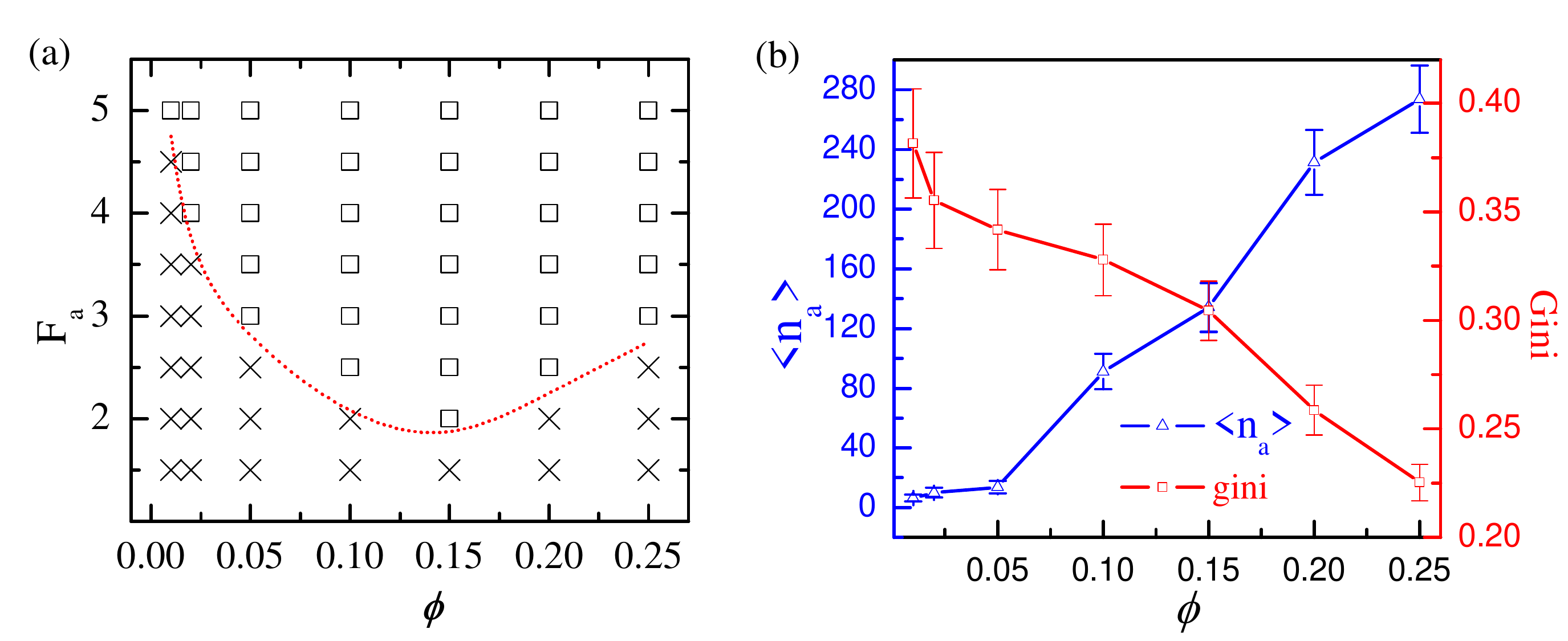}
  \caption{(a) Phase diagram of the G($ \times$)-S($ \Box$) transition in $F_{a}- \phi$ space in the small $D_{r}$ limit ($D_{r}=0.001$ and $N_{p}=100$). The red dash line roughly marks the boundary of the transition. (b) The average number of neighboring particles around the chain and the Gini coefficient of the forces on the boundary beads as a function of particle area fraction, $\phi$ ($F_{a}=1.5$).} \label{fig:4}
\end{figure}

The density of SPPs $\phi$ is another key parameter that influences the G-S and S-G transition. Figure ~\ref{fig:4}(a) shows the G-S transition in $F_a$-$\phi$ space for small $D_r=0.001$. Unexpectedly, $F_{a}^{c}$ is a non-monotonic function of $\phi$. It reaches the minimum around $\phi\approx 0.15$. A possible explanation of such non-monotonic behavior is that the variation of particle density has two opposite impacts on the G-S transition. On the one hand, as the particle density increases, more particles surround the collapsed chain (Fig.~\ref{fig:4}(b)); thus the disturbance and energy transfer to the chain are enhanced. This is the positive side, facilitating the unfolding of the collapsed chain. On the other hand, the distribution of particles surrounding the collapsed chain becomes more homogeneous when the density increases; hence the collisions from the particles turn to be isotropic pressure instead of shear. This is the negative side, impeding the unfolding of the collapsed chain. To quantify the inhomogeneity of the collisions on the chain, we define the Gini coefficient~\cite{Marchetti2014,tian15} as $Gini=\frac{1}{2N_{b}^{2} \overline{|f|}} \sum\limits_{i=1}^{N_{b}} \sum\limits_{j=1}^{N_{b}} ||f_{i}|-|f_{j}||$, where $f_{i}$ is the component force on a boundary bead toward the center of mass of the chain, and $\overline{|f|}$ the mean magnitude of the component forces. The Gini coefficient approaches 0 when the collisions are perfectly homogeneous (pressure effect). The shear effect which causes the deformation and stretch of the chain at small $D_r$ limit will be manifested when the Gini coefficient is large. Figure ~\ref{fig:4}(b) shows that the Gini coefficient decreases with the increase of density. The competition of the above two opposite impacts leads to the non-monotonic dependence of $F_{a}^{c}$ on the particle density. With the parameters in our simulation, the optimal particle density $\phi\approx 0.15$, at which the comprehensive disturbance on the collapsed chain is strongest, i.e. the minimum $F_a$ is required for the G-S transition. We also explore the density dependence of the G-S transition in the large $D_r$ limit (see Fig.S5)~\cite{support}. Since the mechanism for the G-S transition is collision-induced melting instead of shear effect, the above mentioned negative impact as the particle density increases is no longer applicable. For large $D_r=0.3$, $F_{a}^{c}$ turns out to decrease monotonically with particle density.

\textit{Discussion}
The folding and unfolding processes have been well studied in the case of a polymer chain in poor solvent and subjected to a tensile force~\cite{Terentjev2015,Verga2002,Zhulina1991,lai1996}. The phenomenon of hysteresis has also been observed, but the mechanism is due to the nucleation barrier between the globule and stretch states~\cite{Terentjev2015,lai1996}, different from the ''sticky"-particle-induced dynamic barrier in our system. Such hysteresis is not as definite as in our system, i.e. it disappears in the long-chain limit or if the force varies slowly enough~\cite{Terentjev2015}. Both the G-S and S-G transitions are first-order transitions in the tensile-force measurement, no matter there is hysteresis or not. In contrast, ''stable" dumbell-like states in between globule and stretch states are obtained (i.e. the S-G transition is not a first-order transition) when $D_r$ is small (hysteresis is present) in our system. This dumbell-like configuration was also observed in the G-S transition of collapsed polymers in elongational flow fields~\cite{Katz2010}. However, it appears only as a transient and unstable intermediate structure during the process of G-S transition.

Hydrodynamic or the solvent-molecule interactions are the foundation of the G-S transition in shear or elongational flow~\cite{Katz2010,katz06}. Undoubtedly, hydrodynamic interactions, which are ignored in this work, also have a significant impact in our system. But, we believe such impact is most probably quantitative rather than qualitative, since here the G-S and S-G transitions are driven by the collisions between active particles and chain beads, instead of the flow field. Additionally, our model corresponds to a self-attracting polymer with no specific interactions. To study the folding and unfolding processes of bio-polymers in the presence of active particles, we have to consider specific interactions ~\cite{tian03}, which may bring significant differences. For example, the G-S and S-G transitions of DNA-Dps complexes by external force are continuous and the hysteresis is ascribed to the cooperative binding between DNA and Dps~\cite{Abbondanzieri2016}.

\emph{Acknowledgements.}
This work is supported by the National Natural Science Foundation of China (NSFC) Nos. 21474074, 21674078 (W.T.), 21574096, 21774091, 21374073(K.C.), and 91027040 (Y.M.).

\end{spacing}

\end{document}